\documentclass[aps,prl,reprint,twocolumn,groupedaddress,fleqn]{revtex4-1}
\usepackage[english]{babel}
\usepackage{graphicx}
\usepackage{amsmath}
\usepackage{bm}% bold math
\usepackage{setspace}
\usepackage{natbib}
\usepackage{epstopdf}
\begin{document}
\title{Experimental evidence of the Frenkel line in supercritical neon}

\author{C. Prescher$^1$, Yu. D. Fomin$^2$,  V. B. Prakapenka$^3$, J. Stefanski$^1$, K. Trachenko$^4$ and V.V. Brazhkin$^2$}

\affiliation{$^1$Institut f\"ur Geologie und Mineralogie,
Universit\"at zu K\"oln, 50674 K\"oln, Germany}
\affiliation{$^2$Institute for High Pressure Physics, Russian
Academy of Sciences, Troitsk 108840, Moscow, Russia} 
\affiliation{$^3$Center for Advanced Radiation Sources,
University of Chicago, Chicago, Illinois 60637, USA}
\affiliation{$^4$School of Physics and Astronomy, Queen Mary
University of London, Mile End Road, London E1 4NS, United
Kingdom}

\begin{abstract}
Recent research suggests that the supercritical state consists of liquidlike and gaslike states where particle dynamics and key system properties are qualitatively different.
We report experimental evidence of the structural crossover in supercritical neon at pressure and temperature conditions significantly exceeding the critical point values: 250$\,P_c$ and 6.6$\,T_c$.
The experimental results show a crossover of the medium-range order structure evidenced by the change of the structure factor with pressure.
We also observe the crossover of the short-range order structure indicated by changes in the coordination number.
The relative width of the crossover is fairly narrow and is smaller than 10-12 \% in pressure and temperature.
By comparing our experimental results with molecular dynamics simulations, we suggest that the observed crossover can be attributed to the Frenkel line and discuss the relationship between the structural crossover and qualitative changes of dynamical and thermodynamic properties of supercritical matter.
\end{abstract}

\pacs{61.20.Gy, 61.20.Ne, 64.60.Kw} \maketitle

\section{I. Introduction}
Supercritical fluids, the state of matter above the critical
point, are considered uniform in terms of structure and
properties. More recently, experiments have shown that
the dynamic structure factor undergoes
qualitative changes in the supercritical state
\cite{Bencivenga2006,Gorelli2006,Simeoni2010}. To explain this, several mechanisms were
proposed which involved analogues of the liquid-gas transitions
in the supercritical state: the Fisher-Widom line
demarcating different regimes of decay of structural correlations
\cite{Fisher1969,Vega1995, Smiechowski2016}; several versions of percolation lines forming
across conditional bonds or particles \cite{Bernabei2008a,Bernabei2008}; or
''thermodynamic'' continuation of the boiling line such as the
Widom line \cite{Xu2005}.
All these proposed mechanisms were later recognized to have issues which are not currently resolved.
The Fisher-Widom line exists only in a stable fluid for model
low-dimensional systems. The percolation lines are defined in
realistic fluids only under specific conditions. The line of maxima of the
correlation length (the Widom line) and other properties such as
heat capacity, compressibility or thermal expansion depend on the
path in the phase diagram and do not extend far beyond the
critical point \cite{Brazhkin2011a,Brazhkin2011,Fomin2015a,Brazhkin2014}.

In 2012, a new dynamic line in the supercritical region of the
phase diagram was proposed, the Frenkel line
\cite{Brazhkin2012b,Brazhkin2012a,Brazhkin2012,Brazhkin2013}. Crossing the Frenkel line (FL) on
temperature increase (pressure decrease) corresponds to the
qualitative change of particle dynamics, from the combined
oscillatory and diffusive motion as in liquids to purely diffusive
motion as in gases. Simultaneously, high-frequency shear rigidity
disappears (transverse excitations become depleted at all
frequencies \cite{Fomin2016}), bringing the specific heat close to $c_v=2k_{\rm B}$
\cite{Brazhkin2013}. Other important changes take place at the
liquidlike to gaslike crossover at the FL or close to it,
including temperature and pressure dependencies of the sound speed, diffusion coefficient, viscosity and thermal conductivity
\cite{Brazhkin2012,Brazhkin2013}.

Practically, the FL can be located on the basis of presence or
absence of oscillations in the velocity autocorrelation function, the criterion
that coincides with $c_v=2k_{\rm B}$ \cite{Brazhkin2013}. The location of
the FL has been calculated for several systems, including Ar, Fe,
H$_2$O, CO$_2$ and CH$_4$ \cite{Brazhkin2012,Brazhkin2013,Fomin2015,Fomin2014,Yang2015} where it was
established that the temperature at the FL is about 3-5 times
higher than the melting temperature at the same pressure. However,
no experimental study of the Frenkel line was performed.

Importantly, the liquidlike to gaslike crossover at the FL should be accompanied by the {\it structural} crossover of the supercritical fluid. This follows from the dynamical crossover of particle motion and can also be inferred on the basis of the relationship between the structure and the thermodynamic crossover as discussed below in detail. Modelling data suggest the crossover of short-range order at the FL \cite{Bolmatov2013}, yet no experimental evidence supports this prediction. Moreover, it is unclear whether the medium-range order, extensively studied in disordered systems, is sensitive to the FL.

Liquid structure was studied at pressures higher than the critical pressure but often at temperatures close to the melting line. Some of the previous work was aimed at elucidating the structure of supercritical water at a temperature higher than the melting temperature \cite{Soper2000}. The structure of argon was studied at high temperature and pressure \cite{Santoro2008}, albeit at conditions below the FL according to our calculations. As a result, no structural crossover was detected. Another study of argon \cite{Bolmatov2015} addressed the structure at supercritical conditions. Similarly to the previous work \cite{Santoro2008}, the pressure and temperature were below the FL. Although sharp changes of structural parameters were discussed, these changes were not related to crossing the FL but coincided with, and were the result of, sharp changes in the chosen pressure-temperature path \cite{comment}.

\begin{figure*}
	\includegraphics[width=7in]{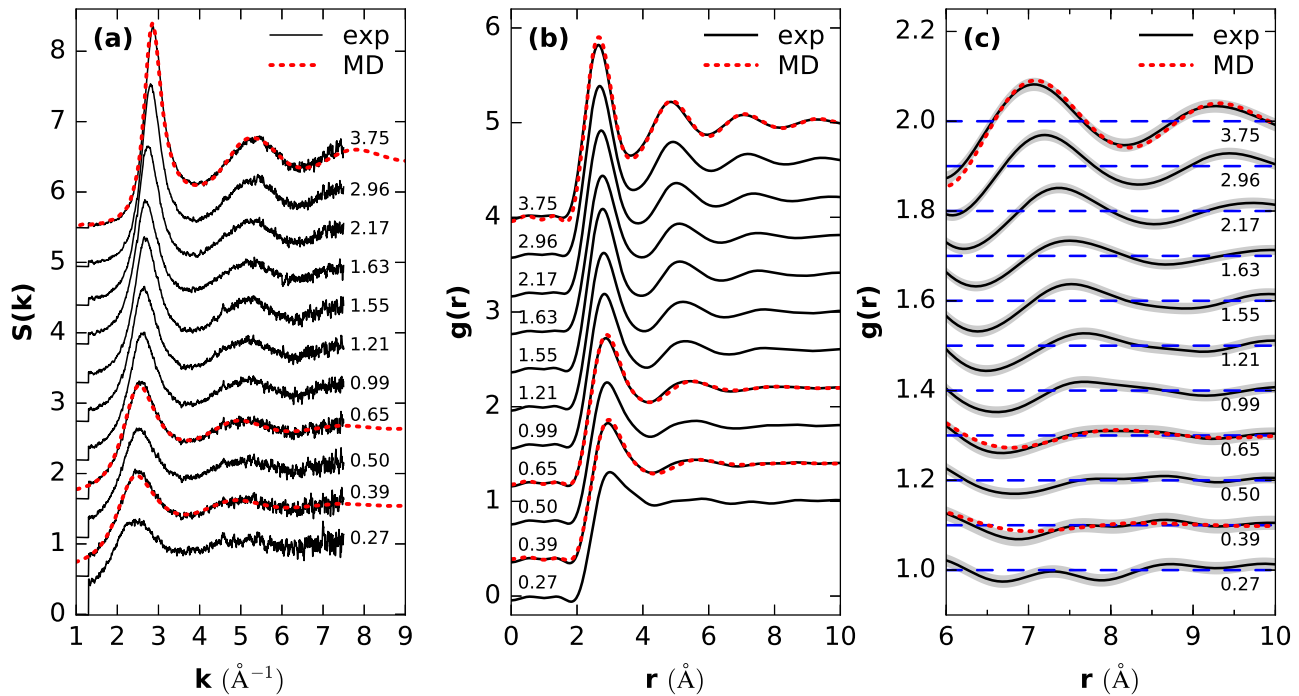}
	\caption{(a) Structure factors $S(k)$ of neon at different pressures.
		(b) Radial distribution functions $g(r)$ of neon obtained by Fourier transform of the structure factors.
		(c) Detail view of $g(r)$ at large distances.
		The pressures are given next to the curves. The black solid and red dashed curves represent experimental (exp) and molecular dynamics (MD) data, respectively. The shaded area in (c) indicates the error, which was calculated by using the standard deviation of $g(r)$ below the first peak; the dashed straight blue line is showing a value of 1 for each $g(r)$ as a guide for the eye. MD radial distribution functions have been calculated with the same cutoff in $S(k)$ as the experimental data and also using a Lorch modification function in order to get comparable results.}
	\label{skgr}
\end{figure*}

Experimental detection of the crossover at the FL in many
interesting systems is challenging, because it requires the
combination of high temperature and relatively low pressure. At
these conditions, many types of standard high-pressure apparatus
such as diamond anvil cells (DAC) can not be used. We have
therefore chosen neon with very low melting and critical
temperature ($T_c=44.5$ K) and fairly high critical pressure
($P_c=2.68$ MPa). Room temperature is 6.6 times higher than
$T_c$, i.e. neon at room temperature and high pressure is a
strongly overheated supercritical fluid. For the Lennard-Jones
(LJ) fluid, the pressure corresponding to $6.6T_c$ at the FL is
about $250P_c$, or $0.6-0.7$ GPa for neon \cite{Brazhkin2013}. This pressure range is
suitable for the DAC.

\section{II. Methods}
The main aim of this study is an experimental detection of a possible structural crossover in supercritical neon at the FL.
We employed symmetric DAC with a culet size of $300\ \mu m$.
Re was used as gasket material and the cell was loaded with Ne at the GSECARS gas-loading system \cite{Rivers2008} to an initial pressure of 0.26 GPa.
The pressure was fine-controlled by membrane system and estimated by the shift of the ruby fluorescence line \cite{Mao1986}.
The x-ray diffraction experiments were performed at the GSECARS, 13-IDD beamline, APS. An incident monochromatic x-ray beam with an energy of 45 keV and $2.5\times3\ \mu m$ spot size was used.
In order to suppress the Compton scattering of the diamond anvils, a multichannel collimator (MCC) as described in Ref. \cite{Weck2013} was employed.
X-ray diffraction data was collected with a Mar345 image plate detector and the geometry was calibrated using LaB$_6$ standard.
Collection time was 300 seconds. The background was measured with an empty cell prior to gas loading.

Detector calibration, image integration and intensity corrections for oblique x-ray to detector angle, cBN seat absorption and diamond absorption were performed using Dioptas software package \cite{Prescher2015a}.
The resulting diffraction patterns were corrected for an additional diamond Compton scattering contribution, which was necessary because the background measurement prior to compression was measured with a thicker sample chamber than the compressed sample at high pressure.
The smaller sample chamber results in more diamond in the volume of diffraction constrained by the MCC. Both the sample signal and the additional diamond Compton scattering contribution were corrected with an MCC transfer function \cite{Weck2013}.

Structure factors and pair distribution functions were calculated following the procedure described in Ref. \cite{Eggert2002}.
We employed the amount of diamond Compton scattering contribution in addition to the density and background scaling as an additional optimization variable.
A Lorch modification function was employed during the Fourier transform in order to minimize unphysical oscillations due to cutoff effects.
Structure factors were obtained up to $k_{\rm max}$=7.5 \AA$^{-1}$.

The experimental study is supported and complemented by molecular
dynamics (MD) simulations. We have used the LJ potential with
parameters $\sigma=2.775$ \AA\ and $\varepsilon=36.831$ K
\cite{Oh2013} and simulated $4000$ particles in a cubic box with
periodic boundary conditions. The equilibration was first
performed in the canonical ensemble at each state point, followed
by the production run in the microcanonical ensemble with $0.2$ fs
timestep for $2 \cdot 10^6$ steps. We find that averaging of calculated properties (such as the speed of sound)
over this large number of timesteps is required to reduce the noise and errors.

We have simulated 15 pressure
points in the range $0.05-3.7$ GPa, corresponding to the density
range of $0.3-2.2$ g/cm$^3$. We have used LAMMPS MD simulation
package \cite{lammps}. The structure factors $S(k)$ were calculated in MD simulations as Fourier transforms of the pair distribution functions $g(r)$:

\begin{equation}
S(k)=1+4 \pi \frac{N}{V} \int_0 ^{\infty} \left( g(r)-1 \right)
\frac{r \sin(kr)}{k}dr,
\end{equation}
where $N/V$ is the number density of fluid.

\section{III. Results and Discussion}

\begin{figure*}
	\includegraphics[width=7in]{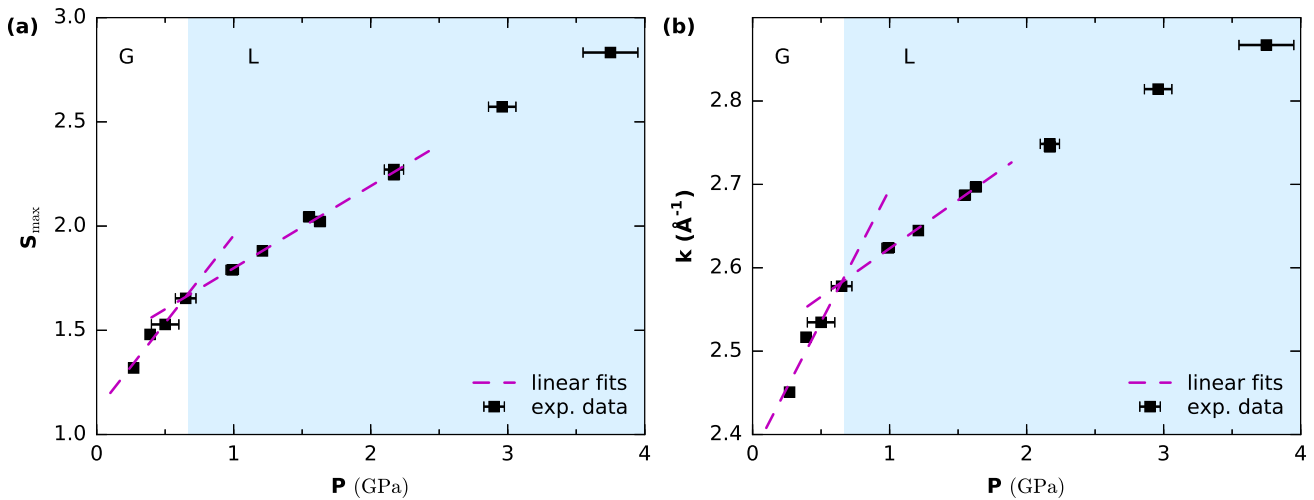}
	
	\caption{
		Extracted height (a) and position (b) of the first maximum of the experimentally derived structure factor $S(k)$ against pressure. 
		The dashed magenta lines show the difference in slopes below and above the crossover.
		The background shadings indicate the region for gaslike (G) and liquidlike (L) supercritical fluid.}
	\label{sk-max}
\end{figure*}

\begin{figure*}
	\includegraphics[width=7in]{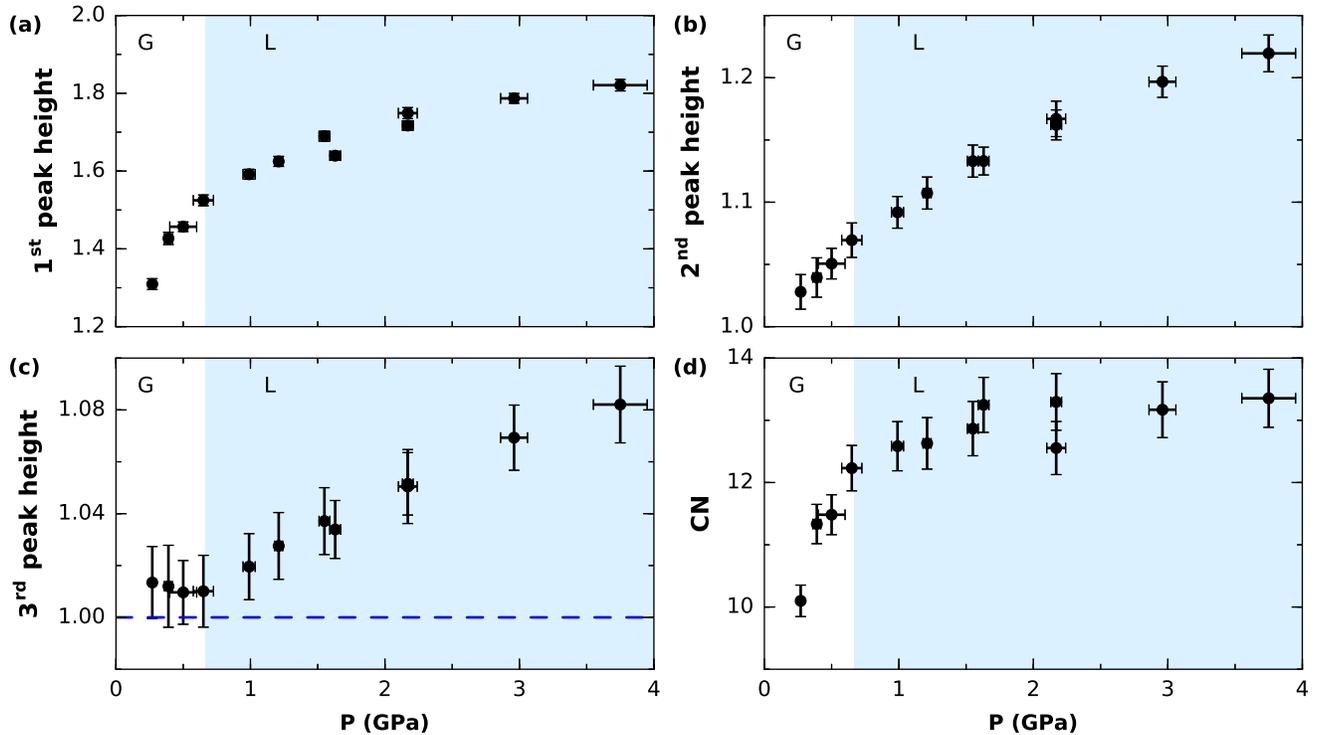}
	\caption{Maxima of the first (a), second (b) and third (c) peak of the experimentally derived $g(r)$ of neon against pressure. Error bars are calculated from the standard deviation of $g(r)$ below the first peak. Dashed blue vertical line in (c) at 1 serves as a guide for the eye. Coordination number (CN) obtained by integrating $g(r)$ up to the minimum after the first peak is shown in (d). The background shadings indicate the region for gaslike (G) and liquidlike (L) supercritical fluid.}
	\label{gr_max}
\end{figure*}

Experimental and simulated structure factors $S(k)$ and their corresponding pair distribution functions $g(r)$ are shown in Fig. \ref{skgr}. We observe very good agreement between experimental results and MD simulations. 
The position and height of the first peak of $S(k)$ (Fig. \ref{sk-max}) show a change of slope against pressure at around $0.65$ GPa. 
We further plot the height of the first three peaks of $g(r)$ vs pressure in Figs. \ref{gr_max}(a-c). While the maxima of the first two peaks show only slight changes of slope with pressure, the maximum of the third peak indicates two regimes: a constant value up to $0.65$ GPa and a linear increase above.
However, the constant height below $0.65$ GPa is only within the error bars [Figs. \ref{skgr}(c), \ref{gr_max}(c)].
Thus, the plateau could be caused by the lack of data accuracy to detect a further decrease in the peak height. 
Nevertheless, the data in Fig. 3(c) show a crossover at 0.65 GPa.

Coordination number (CN) of neon against pressure is shown in Fig. \ref{gr_max}(d). CN was obtained by integration over the first peak of $g(r)$ up to the first minimum $r_{\rm min}$ after the peak using:
\begin{equation}
CN = 4\pi \rho \int_{0}^{r_{\rm min}}r^2 g(r)dr
\end{equation}
\noindent where $\rho=N/V$ is number density.
We observe a strong increase in CN from $10.1$ at $0.27$ GPa to $12.2$ around $0.65$ GPa. At higher pressures CN increases only slightly up to $13.1$ at $3.75$ GPa.
Thus, we observe an increase of a CN with relatively low values, indicating a loosely packed gaslike structure, to a close-packed more liquidlike structure around 0.65 GPa.

\begin{figure*}
	\includegraphics[width=183mm]{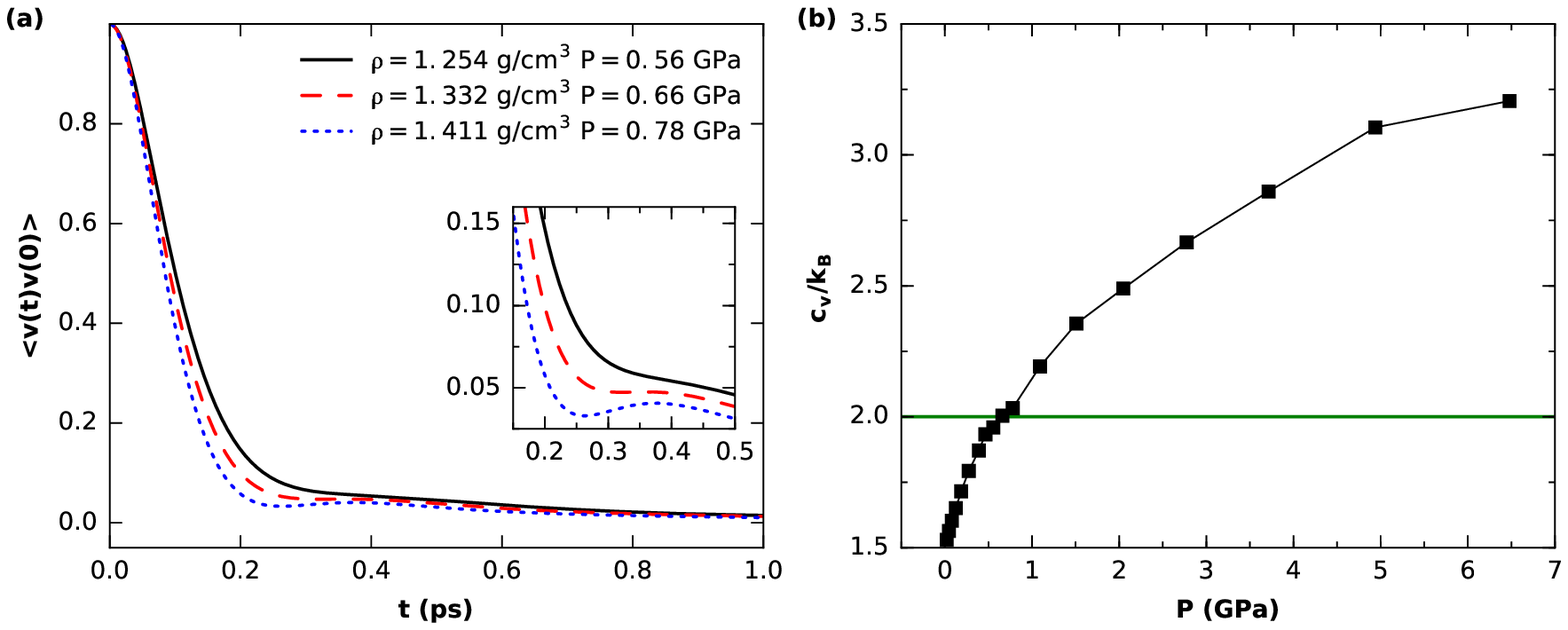}
	\caption{
		(a) Velocity autocorrelation functions of Ne derived from MD simulations in the vicinity of the crossover pressure. (b) Isochoric heat capacities	of neon derived from MD simulations at $T=290$ K. The horizontal line marks $c_v=2k_{\rm B}$.}
	\label{vacf}
\end{figure*}

\begin{figure}
	\includegraphics[]{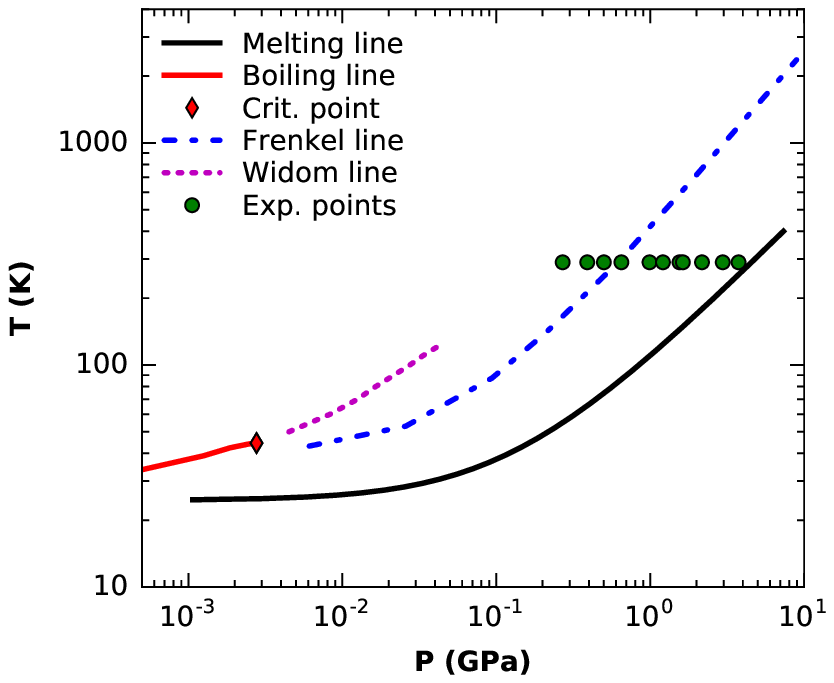}
	\caption{The calculated Frenkel line for neon, shown together with the boiling line, melting line, Widom line (calculated from maximum of heat capacity taken from Ref. \cite{Lemmon}) and experimental points in this study.}
	\label{pd}
\end{figure}

The changes of $S(k)$ and $g(r)$ show a combined modification of the short-range and
medium-range order structure. To relate this crossover to the FL,
we have calculated the position of the FL from MD simulations using two criteria: disappearance of oscillations of velocity
autocorrelation function (VAF) and $c_v=2k_{\rm B}$ \cite{Brazhkin2013}.
VAFs are defined as $Z(t)=\frac{1}{3N} \langle \sum \frac{{\bf
V}_{i}(t) {\bf V}_i(0)}{{\bf V}_i(0)^2} \rangle $ where ${\bf
V}_i(t)$ is $i$-th particle velocity at time $t$. The heat
capacities were obtained from the fluctuations of kinetic energy
in microcanonical ensemble: $<K^2>-<K>^2=\frac{3k_B^2T^2}{2N}(1-\frac{3k_B}{2c_V})$, where $K$
is the kinetic energy of the system \cite{Frenkel2002}. We show the examples of VAF and $c_v$ in Fig. \ref{vacf}. The two criteria
result in almost perfectly coinciding FL curves in Fig. \ref{pd}.

From the calculations according to the two criteria above, we find that at
room temperature the FL is at $0.65\pm 0.02$ GPa, the same pressure
where we have observed the structural crossover in our experiments.

Our data enable us to estimate the width of the crossover at the FL.
The height and position of the first peak of $S(k)$ (see Fig. \ref{sk-max}) as well as the maximum of the third peak undergo a crossover in the range 0.65-0.99 GPa, half-width of which, 0.17 GPa, is usually taken as the maximal width of the crossover.
This gives the pressure at the FL crossover as $P_{\rm F}=0.65\pm 0.08$ GPa. Using the slope of the calculated FL (see below), this gives the crossover temperature $T_{\rm F}=300\pm 30$ K.
Therefore, the relative width of the FL crossover in pressure and temperature is smaller than 10-12\%.

The microscopic origin of the structural crossover at the FL is
related to the qualitative change of particle dynamics. As
discussed above, below the FL particles oscillate around
quasiequilibrium positions and occasionally jump between them.
The average time between jumps is conveniently quantified by
liquid relaxation time $\tau$. This implies that a static
structure exists during $\tau$ for a large number of particles,
giving rise to the well-defined medium-range order comparable to
that existing in structurally disordered solids. 

Therefore, we expect liquidlike structural correlations at distances extending the cage size and beyond, e.g. at larger distances than 6 \AA\ [see Fig. 1c].
Above the FL the particles lose the oscillatory component of motion and start to move in a purely diffusive manner as in gases.
A recent study reporting high-resolution MD simulations using Lennard-Jones potentials of different supercritical fluids \cite{Wang2017} found differences in the temperature dependence of the g(r) peak heights at the FL.
Our experimental data suggests a constant third peak height in the gaslike region indicating loss of medium range order across the FL.
However, the error bars of the experimental results are large and the data might not be accurate enough to observe a further decrease in third peak height or slight deviations from the pressure dependence of the first two peaks in g(r).
Therefore, a clear answer to whether a loss of medium range order or only a change in the pressure dependence of the peak heights occur at the FL cannot be given based on our current data.
The question also arises whether Lennard-Jones potentials can accurately model all interactions of even simple supercritical fluids like neon.
While we observe a good agreement between our experimental and simulation data, the agreement is still not absolutely perfect.
The interaction might need to be modeled by more sophisticated potentials or even with first-principle methods.
A more clear explanation for the change of particle dynamics at the FL comes from the change of CN. 
The different dynamical regimes are characterized by the change from a close-packed
local configuration with 12-fold and above coordination below the FL to a less densely packed above the FL [see Fig. \ref{gr_max}(d)].

Another interesting insight into the origin of the structural
crossover comes from the relationship between structure and
thermodynamics. The system energy can be written as an integral
over the pair distribution function $g(r)$ as

\begin{equation}
E=\frac{3}{2}k_{\rm B}T+2 \pi \rho \int\limits_0^{\infty} r^2
U(r)g(r)dr, \label{ene}
\end{equation}
\noindent where $\rho=N/V$ is number density and $U(r)$ is interatomic potential.

For the case of harmonic modes, phonons, $E$ in (2) can be written as $E=E_0+E_T$, where $E_0$ is the energy at zero temperature and $E_T$ is the phonon thermal energy.
The collective modes undergo the crossover at the FL,
and so does the energy. In particular, the transverse modes below
the FL start disappearing starting from the lowest frequency equal
$\frac{1}{\tau}$ \cite{Trachenko2016} and disappear completely above the FL \cite{Fomin2016}.
Above the FL, the remaining longitudinal mode starts disappearing
starting from the highest frequency $\frac{2\pi c}{L}$, where $L$
is the particle mean free path (no oscillations can take place at
distance smaller than $L$) \cite{Trachenko2016}. This gives qualitatively
different behavior of the energy below and above the FL, resulting
in the crossover at the FL. According to (\ref{ene}), the
crossover of energy necessarily implies the crossover of $g(r)$.

We have confirmed this mechanism by calculating the collective
modes and their dispersion curves directly. We calculate the
longitudinal and transverse current correlation functions $C_L$
and $C_T$ as

\begin{equation}
\begin{aligned}
&C_L(k,t)=\frac{k^2}{N}\langle J_z({\bf k},t) \cdot J_z(-{\bf
k},0)\rangle\\
&C_T(k,t)=\frac{k^2}{2N} \langle J_x({\bf k},t)\cdot J_x(-{\bf
k},0)+J_y({\bf k},t) \cdot J_y(-{\bf k},0)\rangle\nonumber
\label{zero}
\end{aligned}
\end{equation}
\noindent where $J({\bf k},t)=\sum_{j=1}^N {\bf v}_j e^{-i{\bf k
r}_j(t)}$ is the velocity current \cite{Hansen2013,Rapaport1995}.

\begin{figure}
	\includegraphics[width=84mm]{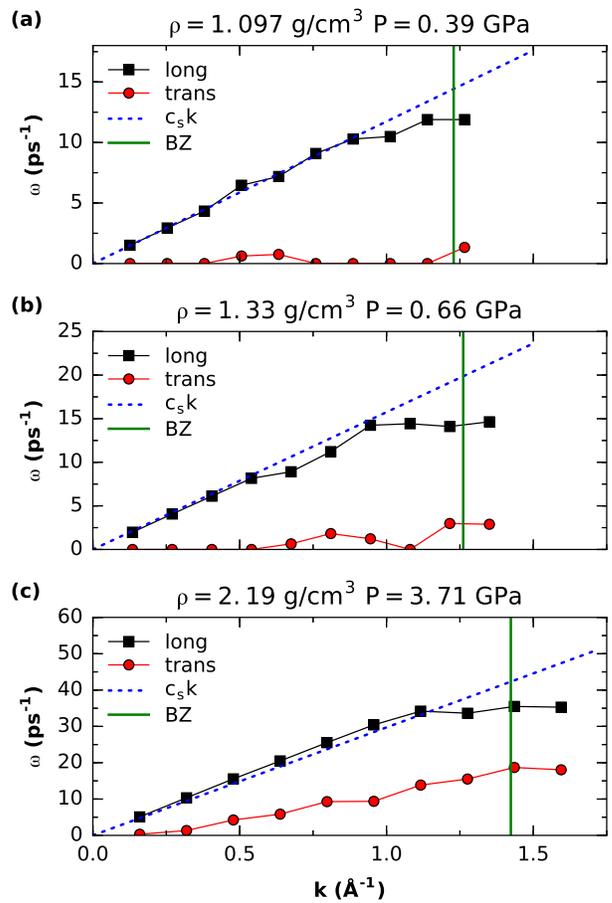}
	\caption{
		Excitation spectra of neon derived from MD simulations at pressure points (a) above (b) nearly at and (c) below the Frenkel line.
		Longitudinal (squares) and transverse (circles) spectra are shown. The dashed lines give the Debye dispersion law $\omega = c_s k$, where $c_s$ is the adiabatic speed of sound.
		The vertical lines mark the boundary of pseudo-Brillouin zone (BZ).}
	\label{sp}
\end{figure}

The maxima of Fourier transforms $\tilde{C}_L({\bf k},\omega)$ and
$\tilde{C}_T({\bf k},\omega)$ give the frequencies of longitudinal
and transverse excitations. The resulting dispersion curves are
shown in Figs. \ref{sp} (a)-(c) at pressures below, nearly at and
above the Frenkel line. We observe that transverse excitations are
seen in a large part of the first pseudo-Brillouin zone at high
pressure below the FL. At low pressure above the line, the
transverse excitations disappear. At the line itself, only small
traces of transverse modes close to the boundary can be resolved.

We therefore find that collective excitations undergo a
qualitative crossover at the FL, consistent with the earlier
predictions as well as with the structural crossover via Eq.
(\ref{ene}).

An interesting consequence of the existence of transverse modes
below the FL is positive sound dispersion (PSD), the increase of
the speed of sound above its adiabatic hydrodynamic value. In
Fig. \ref{sp} we show the adiabatic speed of sound calculated as
$c_s=\gamma ^{1/2} c_T$, where $c_T=\left ( \frac{dP}{d \rho}
\right )^{1/2}_T$ is the isothermal speed of sound and $\gamma =
c_P/c_V$ is the ratio of isobaric and isochoric heat capacities.
The isobaric heat capacity was calculated as
$c_P=c_V+\frac{T}{\rho ^2}\left(\frac{\partial P}{\partial T}
\right)^2_{\rho} \left(\frac{\partial P}{\partial \rho}
\right)^{-1}_T$.

We observe the PSD below the FL but not above. This is consistent with the known effect of increase
of the longitudinal speed of sound in the presence of shear rigidity (shear waves) over its
hydrodynamic value \cite{Trachenko2016,Fomin2016}. Therefore, the PSD can be discussed on the basis
of a visco-elastic model as a result of the presence of transverse modes in the supercritical system.
We note that PSD was previously related to the Widom line \cite{Xu2005} in view of the change of the
dynamic structure factor \cite{Bencivenga2006,Gorelli2006,Simeoni2010}. However, we
note that the Widom line exists close to the critical point only and disappears well above the critical
point as is the case for conditions discussed in this work \cite{Brazhkin2011a,Brazhkin2011,Fomin2015a,Brazhkin2014}.

Another interesting consequence of our results is the possibility to
observe liquid-liquid phase transitions in the supercritical
state. So far liquid-liquid transitions involving the change of
the medium-range order have been observed below the critical point
only \cite{Brazhkin2002}. In this work, we have ascertained that the
medium-range is present in the supercritical state too, as long as
the system is below the FL. Therefore, we propose that
liquid-liquid phase transitions can be observed above the critical
point below the FL.

We propose that our results can be relevant for industrial
application of supercritical technologies \cite{Kiran2000}.
Supercritical fluids combine high density ($2-3$ times higher than
the density at the triple point) and high diffusion coefficient
that are orders of magnitude higher than those of subcritical
liquids. This leads to remarkable increase of solubility and
speedup of chemical reactions. Interestingly, the solubility
maxima lie very close to the Frenkel line \cite{Yang2015}
increasing pressure along the FL gives higher density and
diffusivity and minimal viscosity. The data of the FL can enhance
the supercritical technologies, particularly at higher pressure in
the range $1-3$ GPa where high-pressure chambers can have large
volume.

In summary, we have directly ascertained a structural crossover in the supercritical state.
Of particular importance is that the crossover operates at extremely high pressure and temperature: $6.6T_c$ and $250P_c$.
The relative width of the crossover is fairly  narrow and is less than 10 \% in both pressure and temperature.
Comparing our experimental results with MD simulations enables us to consider the Frenkel line as a boundary between liquidlike and gaslike states of supercritical matter.

\section{Acknowledgements}
We are grateful to G. Simeoni, V. N. Ryzhov, E. N. Tsiok and J. E.
Proctor for fruitful discussions. Yu.D.F. thanks the Russian
Scientific Center at Kurchatov Institute and Joint Supercomputing
Center of Russian Academy of Science for computational facilities.
Yu.D.F. (molecular dynamic simulations, the discussions of the results, writing paper) and V.V.B. (formulation of the problem, the discussions of the results, writing paper) are grateful to Russian Science Foundation  (14-22-00093) for the financial support.
K.T. is grateful to the Royal Society for support.
We thank S. Tkachev for his help during gas-loading.
Use of the COMPRES-GSECARS gas loading system was supported by COMPRES under NSF Cooperative Agreement EAR 11-57758 and by GSECARS through NSF Grant No. EAR-1128799 and DOE Grant No. DE-FG02-94ER14466.
Portions of this work were performed at GeoSoilEnviroCARS The University of Chicago, Sector 13), Advanced Photon Source (APS), Argonne National Laboratory.
GeoSoilEnviroCARS is supported by the National Science Foundation - Earth Sciences (EAR-1128799) and Department of Energy- GeoSciences (DE-FG02-94ER14466).
This research used resources of the Advanced Photon Source, a U.S. Department of Energy Office of Science User Facility operated for the DOE Office of Science by Argonne National Laboratory under Contract No. DE-AC02-06CH11357.

\bibliographystyle{apsrev4-1}
\bibliography{Mendeley.bib}% Produces the bibliography via BibTeX.

\begin{thebibliography}{41}%
\makeatletter
\providecommand \@ifxundefined [1]{%
 \@ifx{#1\undefined}
}%
\providecommand \@ifnum [1]{%
 \ifnum #1\expandafter \@firstoftwo
 \else \expandafter \@secondoftwo
 \fi
}%
\providecommand \@ifx [1]{%
 \ifx #1\expandafter \@firstoftwo
 \else \expandafter \@secondoftwo
 \fi
}%
\providecommand \natexlab [1]{#1}%
\providecommand \enquote  [1]{``#1''}%
\providecommand \bibnamefont  [1]{#1}%
\providecommand \bibfnamefont [1]{#1}%
\providecommand \citenamefont [1]{#1}%
\providecommand \href@noop [0]{\@secondoftwo}%
\providecommand \href [0]{\begingroup \@sanitize@url \@href}%
\providecommand \@href[1]{\@@startlink{#1}\@@href}%
\providecommand \@@href[1]{\endgroup#1\@@endlink}%
\providecommand \@sanitize@url [0]{\catcode `\\12\catcode `\$12\catcode
  `\&12\catcode `\#12\catcode `\^12\catcode `\_12\catcode `\%12\relax}%
\providecommand \@@startlink[1]{}%
\providecommand \@@endlink[0]{}%
\providecommand \url  [0]{\begingroup\@sanitize@url \@url }%
\providecommand \@url [1]{\endgroup\@href {#1}{\urlprefix }}%
\providecommand \urlprefix  [0]{URL }%
\providecommand \Eprint [0]{\href }%
\providecommand \doibase [0]{http://dx.doi.org/}%
\providecommand \selectlanguage [0]{\@gobble}%
\providecommand \bibinfo  [0]{\@secondoftwo}%
\providecommand \bibfield  [0]{\@secondoftwo}%
\providecommand \translation [1]{[#1]}%
\providecommand \BibitemOpen [0]{}%
\providecommand \bibitemStop [0]{}%
\providecommand \bibitemNoStop [0]{.\EOS\space}%
\providecommand \EOS [0]{\spacefactor3000\relax}%
\providecommand \BibitemShut  [1]{\csname bibitem#1\endcsname}%
\let\auto@bib@innerbib\@empty
%</preamble>
\bibitem [{\citenamefont {Bencivenga}\ \emph {et~al.}(2006)\citenamefont
  {Bencivenga}, \citenamefont {Cunsolo}, \citenamefont {Krisch}, \citenamefont
  {Monaco}, \citenamefont {Ruocco},\ and\ \citenamefont
  {Sette}}]{Bencivenga2006}%
  \BibitemOpen
  \bibfield  {author} {\bibinfo {author} {\bibfnamefont {F.}~\bibnamefont
  {Bencivenga}}, \bibinfo {author} {\bibfnamefont {A.}~\bibnamefont {Cunsolo}},
  \bibinfo {author} {\bibfnamefont {M.}~\bibnamefont {Krisch}}, \bibinfo
  {author} {\bibfnamefont {G.}~\bibnamefont {Monaco}}, \bibinfo {author}
  {\bibfnamefont {G.}~\bibnamefont {Ruocco}}, \ and\ \bibinfo {author}
  {\bibfnamefont {F.}~\bibnamefont {Sette}},\ }\href {\doibase
  10.1209/epl/i2006-10091-y} {\bibfield  {journal} {\bibinfo  {journal}
  {Europhys. Lett.}\ }\textbf {\bibinfo {volume} {75}},\ \bibinfo {pages} {70}
  (\bibinfo {year} {2006})}\BibitemShut {NoStop}%
\bibitem [{\citenamefont {Gorelli}\ \emph {et~al.}(2006)\citenamefont
  {Gorelli}, \citenamefont {Santoro}, \citenamefont {Scopigno}, \citenamefont
  {Krisch},\ and\ \citenamefont {Ruocco}}]{Gorelli2006}%
  \BibitemOpen
  \bibfield  {author} {\bibinfo {author} {\bibfnamefont {F.}~\bibnamefont
  {Gorelli}}, \bibinfo {author} {\bibfnamefont {M.}~\bibnamefont {Santoro}},
  \bibinfo {author} {\bibfnamefont {T.}~\bibnamefont {Scopigno}}, \bibinfo
  {author} {\bibfnamefont {M.}~\bibnamefont {Krisch}}, \ and\ \bibinfo {author}
  {\bibfnamefont {G.}~\bibnamefont {Ruocco}},\ }\href {\doibase
  10.1103/PhysRevLett.97.245702} {\bibfield  {journal} {\bibinfo  {journal}
  {Phys. Rev. Lett.}\ }\textbf {\bibinfo {volume} {97}},\ \bibinfo {pages}
  {245702} (\bibinfo {year} {2006})}\BibitemShut {NoStop}%
\bibitem [{\citenamefont {Simeoni}\ \emph {et~al.}(2010)\citenamefont
  {Simeoni}, \citenamefont {Bryk}, \citenamefont {Gorelli}, \citenamefont
  {Krisch}, \citenamefont {Ruocco}, \citenamefont {Santoro},\ and\
  \citenamefont {Scopigno}}]{Simeoni2010}%
  \BibitemOpen
  \bibfield  {author} {\bibinfo {author} {\bibfnamefont {G.~G.}\ \bibnamefont
  {Simeoni}}, \bibinfo {author} {\bibfnamefont {T.}~\bibnamefont {Bryk}},
  \bibinfo {author} {\bibfnamefont {F.~a.}\ \bibnamefont {Gorelli}}, \bibinfo
  {author} {\bibfnamefont {M.}~\bibnamefont {Krisch}}, \bibinfo {author}
  {\bibfnamefont {G.}~\bibnamefont {Ruocco}}, \bibinfo {author} {\bibfnamefont
  {M.}~\bibnamefont {Santoro}}, \ and\ \bibinfo {author} {\bibfnamefont
  {T.}~\bibnamefont {Scopigno}},\ }\href {\doibase 10.1038/nphys1683}
  {\bibfield  {journal} {\bibinfo  {journal} {Nature Phys.}\ }\textbf {\bibinfo
  {volume} {6}},\ \bibinfo {pages} {503} (\bibinfo {year} {2010})}\BibitemShut
  {NoStop}%
\bibitem [{\citenamefont {Fisher}\ and\ \citenamefont
  {Widom}(1969)}]{Fisher1969}%
  \BibitemOpen
  \bibfield  {author} {\bibinfo {author} {\bibfnamefont {M.~E.}\ \bibnamefont
  {Fisher}}\ and\ \bibinfo {author} {\bibfnamefont {B.}~\bibnamefont {Widom}},\
  }\href {\doibase 10.1063/1.1671624} {\bibfield  {journal} {\bibinfo
  {journal} {J. Chem. Phys.}\ }\textbf {\bibinfo {volume} {50}},\ \bibinfo
  {pages} {3756} (\bibinfo {year} {1969})}\BibitemShut {NoStop}%
\bibitem [{\citenamefont {Vega}\ \emph {et~al.}(1995)\citenamefont {Vega},
  \citenamefont {Rull},\ and\ \citenamefont {Lago}}]{Vega1995}%
  \BibitemOpen
  \bibfield  {author} {\bibinfo {author} {\bibfnamefont {C.}~\bibnamefont
  {Vega}}, \bibinfo {author} {\bibfnamefont {L.~F.}\ \bibnamefont {Rull}}, \
  and\ \bibinfo {author} {\bibfnamefont {S.}~\bibnamefont {Lago}},\ }\href
  {\doibase 10.1103/PhysRevE.51.3146} {\bibfield  {journal} {\bibinfo
  {journal} {Phys. Rev. E}\ }\textbf {\bibinfo {volume} {51}},\ \bibinfo
  {pages} {3146} (\bibinfo {year} {1995})}\BibitemShut {NoStop}%
\bibitem [{\citenamefont {Smiechowski}\ \emph {et~al.}(2016)\citenamefont
  {Smiechowski}, \citenamefont {Schran}, \citenamefont {Forbert},\ and\
  \citenamefont {Marx}}]{Smiechowski2016}%
  \BibitemOpen
  \bibfield  {author} {\bibinfo {author} {\bibfnamefont {M.}~\bibnamefont
  {Smiechowski}}, \bibinfo {author} {\bibfnamefont {C.}~\bibnamefont {Schran}},
  \bibinfo {author} {\bibfnamefont {H.}~\bibnamefont {Forbert}}, \ and\
  \bibinfo {author} {\bibfnamefont {D.}~\bibnamefont {Marx}},\ }\href {\doibase
  10.1103/PhysRevLett.116.027801} {\bibfield  {journal} {\bibinfo  {journal}
  {Phys. Rev. Lett.}\ }\textbf {\bibinfo {volume} {116}},\ \bibinfo {pages}
  {027801} (\bibinfo {year} {2016})}\BibitemShut {NoStop}%
\bibitem [{\citenamefont {Bernabei}\ \emph {et~al.}(2008)\citenamefont
  {Bernabei}, \citenamefont {Botti}, \citenamefont {Bruni}, \citenamefont
  {Ricci},\ and\ \citenamefont {Soper}}]{Bernabei2008a}%
  \BibitemOpen
  \bibfield  {author} {\bibinfo {author} {\bibfnamefont {M.}~\bibnamefont
  {Bernabei}}, \bibinfo {author} {\bibfnamefont {A.}~\bibnamefont {Botti}},
  \bibinfo {author} {\bibfnamefont {F.}~\bibnamefont {Bruni}}, \bibinfo
  {author} {\bibfnamefont {M.~A.}\ \bibnamefont {Ricci}}, \ and\ \bibinfo
  {author} {\bibfnamefont {A.~K.}\ \bibnamefont {Soper}},\ }\href {\doibase
  10.1103/PhysRevE.78.021505} {\bibfield  {journal} {\bibinfo  {journal} {Phys.
  Rev. E}\ }\textbf {\bibinfo {volume} {78}},\ \bibinfo {pages} {021505}
  (\bibinfo {year} {2008})}\BibitemShut {NoStop}%
\bibitem [{\citenamefont {Bernabei}\ and\ \citenamefont
  {Ricci}(2008)}]{Bernabei2008}%
  \BibitemOpen
  \bibfield  {author} {\bibinfo {author} {\bibfnamefont {M.}~\bibnamefont
  {Bernabei}}\ and\ \bibinfo {author} {\bibfnamefont {M.~A.}\ \bibnamefont
  {Ricci}},\ }\href {\doibase 10.1088/0953-8984/20/49/494208} {\bibfield
  {journal} {\bibinfo  {journal} {J. Phys. Cond. Matt.}\ }\textbf {\bibinfo
  {volume} {20}},\ \bibinfo {pages} {494208} (\bibinfo {year}
  {2008})}\BibitemShut {NoStop}%
\bibitem [{\citenamefont {Xu}\ \emph {et~al.}(2005)\citenamefont {Xu},
  \citenamefont {Kumar}, \citenamefont {Buldyrev}, \citenamefont {Chen},
  \citenamefont {Poole}, \citenamefont {Sciortino},\ and\ \citenamefont
  {Stanley}}]{Xu2005}%
  \BibitemOpen
  \bibfield  {author} {\bibinfo {author} {\bibfnamefont {L.}~\bibnamefont
  {Xu}}, \bibinfo {author} {\bibfnamefont {P.}~\bibnamefont {Kumar}}, \bibinfo
  {author} {\bibfnamefont {S.~V.}\ \bibnamefont {Buldyrev}}, \bibinfo {author}
  {\bibfnamefont {S.-H.}\ \bibnamefont {Chen}}, \bibinfo {author}
  {\bibfnamefont {P.~H.}\ \bibnamefont {Poole}}, \bibinfo {author}
  {\bibfnamefont {F.}~\bibnamefont {Sciortino}}, \ and\ \bibinfo {author}
  {\bibfnamefont {H.~E.}\ \bibnamefont {Stanley}},\ }\href {\doibase
  10.1073/pnas.0507870102} {\bibfield  {journal} {\bibinfo  {journal} {Proc.
  Natl. Acad. Sci. U. S. A.}\ }\textbf {\bibinfo {volume} {102}},\ \bibinfo
  {pages} {16558} (\bibinfo {year} {2005})}\BibitemShut {NoStop}%
\bibitem [{\citenamefont {Brazhkin}\ \emph
  {et~al.}(2011{\natexlab{a}})\citenamefont {Brazhkin}, \citenamefont {Fomin},
  \citenamefont {Lyapin}, \citenamefont {Ryzhov},\ and\ \citenamefont
  {Tsiok}}]{Brazhkin2011a}%
  \BibitemOpen
  \bibfield  {author} {\bibinfo {author} {\bibfnamefont {V.~V.}\ \bibnamefont
  {Brazhkin}}, \bibinfo {author} {\bibfnamefont {Y.~D.}\ \bibnamefont {Fomin}},
  \bibinfo {author} {\bibfnamefont {A.~G.}\ \bibnamefont {Lyapin}}, \bibinfo
  {author} {\bibfnamefont {V.~N.}\ \bibnamefont {Ryzhov}}, \ and\ \bibinfo
  {author} {\bibfnamefont {E.~N.}\ \bibnamefont {Tsiok}},\ }\href {\doibase
  10.1021/jp2039898} {\bibfield  {journal} {\bibinfo  {journal} {J. Phys. Chem.
  B}\ }\textbf {\bibinfo {volume} {115}},\ \bibinfo {pages} {14112} (\bibinfo
  {year} {2011}{\natexlab{a}})}\BibitemShut {NoStop}%
\bibitem [{\citenamefont {Brazhkin}\ \emph
  {et~al.}(2011{\natexlab{b}})\citenamefont {Brazhkin}, \citenamefont
  {Lyapin},\ and\ \citenamefont {Trachenko}}]{Brazhkin2011}%
  \BibitemOpen
  \bibfield  {author} {\bibinfo {author} {\bibfnamefont {V.~V.}\ \bibnamefont
  {Brazhkin}}, \bibinfo {author} {\bibfnamefont {A.~G.}\ \bibnamefont
  {Lyapin}}, \ and\ \bibinfo {author} {\bibfnamefont {K.}~\bibnamefont
  {Trachenko}},\ }\href {\doibase 10.1103/PhysRevB.83.132103} {\bibfield
  {journal} {\bibinfo  {journal} {Physical Review B}\ }\textbf {\bibinfo
  {volume} {83}},\ \bibinfo {pages} {2} (\bibinfo {year}
  {2011}{\natexlab{b}})}\BibitemShut {NoStop}%
\bibitem [{\citenamefont {Fomin}\ \emph
  {et~al.}(2015{\natexlab{a}})\citenamefont {Fomin}, \citenamefont {Ryzhov},
  \citenamefont {Tsiok},\ and\ \citenamefont {Brazhkin}}]{Fomin2015a}%
  \BibitemOpen
  \bibfield  {author} {\bibinfo {author} {\bibfnamefont {Y.~D.}\ \bibnamefont
  {Fomin}}, \bibinfo {author} {\bibfnamefont {V.~N.}\ \bibnamefont {Ryzhov}},
  \bibinfo {author} {\bibfnamefont {E.~N.}\ \bibnamefont {Tsiok}}, \ and\
  \bibinfo {author} {\bibfnamefont {V.~V.}\ \bibnamefont {Brazhkin}},\ }\href
  {\doibase 10.1103/PhysRevE.91.022111} {\bibfield  {journal} {\bibinfo
  {journal} {Phys. Rev. E}\ }\textbf {\bibinfo {volume} {91}},\ \bibinfo
  {pages} {022111} (\bibinfo {year} {2015}{\natexlab{a}})}\BibitemShut
  {NoStop}%
\bibitem [{\citenamefont {Brazhkin}\ \emph {et~al.}(2014)\citenamefont
  {Brazhkin}, \citenamefont {Fomin}, \citenamefont {Ryzhov}, \citenamefont
  {Tareyeva},\ and\ \citenamefont {Tsiok}}]{Brazhkin2014}%
  \BibitemOpen
  \bibfield  {author} {\bibinfo {author} {\bibfnamefont {V.~V.}\ \bibnamefont
  {Brazhkin}}, \bibinfo {author} {\bibfnamefont {Y.~D.}\ \bibnamefont {Fomin}},
  \bibinfo {author} {\bibfnamefont {V.~N.}\ \bibnamefont {Ryzhov}}, \bibinfo
  {author} {\bibfnamefont {E.~E.}\ \bibnamefont {Tareyeva}}, \ and\ \bibinfo
  {author} {\bibfnamefont {E.~N.}\ \bibnamefont {Tsiok}},\ }\href {\doibase
  10.1103/PhysRevE.89.042136} {\bibfield  {journal} {\bibinfo  {journal} {Phys.
  Rev. E}\ }\textbf {\bibinfo {volume} {89}},\ \bibinfo {pages} {042136}
  (\bibinfo {year} {2014})}\BibitemShut {NoStop}%
\bibitem [{\citenamefont {Brazhkin}\ \emph
  {et~al.}(2012{\natexlab{a}})\citenamefont {Brazhkin}, \citenamefont {Lyapin},
  \citenamefont {Ryzhov}, \citenamefont {Trachenko}, \citenamefont {Fomin},\
  and\ \citenamefont {Tsiok}}]{Brazhkin2012b}%
  \BibitemOpen
  \bibfield  {author} {\bibinfo {author} {\bibfnamefont {V.~V.}\ \bibnamefont
  {Brazhkin}}, \bibinfo {author} {\bibfnamefont {A.~G.}\ \bibnamefont
  {Lyapin}}, \bibinfo {author} {\bibfnamefont {V.~N.}\ \bibnamefont {Ryzhov}},
  \bibinfo {author} {\bibfnamefont {K.}~\bibnamefont {Trachenko}}, \bibinfo
  {author} {\bibfnamefont {Y.~D.}\ \bibnamefont {Fomin}}, \ and\ \bibinfo
  {author} {\bibfnamefont {E.~N.}\ \bibnamefont {Tsiok}},\ }\href {\doibase
  10.3367/UFNe.0182.201211a.1137} {\bibfield  {journal} {\bibinfo  {journal}
  {Physics-Uspekhi}\ }\textbf {\bibinfo {volume} {55}},\ \bibinfo {pages}
  {1061} (\bibinfo {year} {2012}{\natexlab{a}})}\BibitemShut {NoStop}%
\bibitem [{\citenamefont {Brazhkin}\ \emph
  {et~al.}(2012{\natexlab{b}})\citenamefont {Brazhkin}, \citenamefont {Fomin},
  \citenamefont {Lyapin}, \citenamefont {Ryzhov},\ and\ \citenamefont
  {Trachenko}}]{Brazhkin2012a}%
  \BibitemOpen
  \bibfield  {author} {\bibinfo {author} {\bibfnamefont {V.~V.}\ \bibnamefont
  {Brazhkin}}, \bibinfo {author} {\bibfnamefont {Y.~D.}\ \bibnamefont {Fomin}},
  \bibinfo {author} {\bibfnamefont {a.~G.}\ \bibnamefont {Lyapin}}, \bibinfo
  {author} {\bibfnamefont {V.~N.}\ \bibnamefont {Ryzhov}}, \ and\ \bibinfo
  {author} {\bibfnamefont {K.}~\bibnamefont {Trachenko}},\ }\href {\doibase
  10.1134/S0021364012030034} {\bibfield  {journal} {\bibinfo  {journal} {JETP
  Lett.}\ }\textbf {\bibinfo {volume} {95}},\ \bibinfo {pages} {164} (\bibinfo
  {year} {2012}{\natexlab{b}})}\BibitemShut {NoStop}%
\bibitem [{\citenamefont {Brazhkin}\ \emph
  {et~al.}(2012{\natexlab{c}})\citenamefont {Brazhkin}, \citenamefont {Fomin},
  \citenamefont {Lyapin}, \citenamefont {Ryzhov},\ and\ \citenamefont
  {Trachenko}}]{Brazhkin2012}%
  \BibitemOpen
  \bibfield  {author} {\bibinfo {author} {\bibfnamefont {V.~V.}\ \bibnamefont
  {Brazhkin}}, \bibinfo {author} {\bibfnamefont {Y.~D.}\ \bibnamefont {Fomin}},
  \bibinfo {author} {\bibfnamefont {A.~G.}\ \bibnamefont {Lyapin}}, \bibinfo
  {author} {\bibfnamefont {V.~N.}\ \bibnamefont {Ryzhov}}, \ and\ \bibinfo
  {author} {\bibfnamefont {K.}~\bibnamefont {Trachenko}},\ }\href {\doibase
  10.1103/PhysRevE.85.031203} {\bibfield  {journal} {\bibinfo  {journal} {Phys.
  Rev. E}\ }\textbf {\bibinfo {volume} {85}},\ \bibinfo {pages} {032103}
  (\bibinfo {year} {2012}{\natexlab{c}})}\BibitemShut {NoStop}%
\bibitem [{\citenamefont {Brazhkin}\ \emph {et~al.}(2013)\citenamefont
  {Brazhkin}, \citenamefont {Fomin}, \citenamefont {Lyapin}, \citenamefont
  {Ryzhov}, \citenamefont {Tsiok},\ and\ \citenamefont
  {Trachenko}}]{Brazhkin2013}%
  \BibitemOpen
  \bibfield  {author} {\bibinfo {author} {\bibfnamefont {V.~V.}\ \bibnamefont
  {Brazhkin}}, \bibinfo {author} {\bibfnamefont {Y.~D.}\ \bibnamefont {Fomin}},
  \bibinfo {author} {\bibfnamefont {A.~G.}\ \bibnamefont {Lyapin}}, \bibinfo
  {author} {\bibfnamefont {V.~N.}\ \bibnamefont {Ryzhov}}, \bibinfo {author}
  {\bibfnamefont {E.~N.}\ \bibnamefont {Tsiok}}, \ and\ \bibinfo {author}
  {\bibfnamefont {K.}~\bibnamefont {Trachenko}},\ }\href {\doibase
  10.1103/PhysRevLett.111.145901} {\bibfield  {journal} {\bibinfo  {journal}
  {Phys. Rev. Lett.}\ }\textbf {\bibinfo {volume} {111}},\ \bibinfo {pages}
  {145901} (\bibinfo {year} {2013})}\BibitemShut {NoStop}%
\bibitem [{\citenamefont {Fomin}\ \emph {et~al.}(2016)\citenamefont {Fomin},
  \citenamefont {Ryzhov}, \citenamefont {Tsiok}, \citenamefont {Brazhkin},\
  and\ \citenamefont {Trachenko}}]{Fomin2016}%
  \BibitemOpen
  \bibfield  {author} {\bibinfo {author} {\bibfnamefont {Y.~D.}\ \bibnamefont
  {Fomin}}, \bibinfo {author} {\bibfnamefont {V.~N.}\ \bibnamefont {Ryzhov}},
  \bibinfo {author} {\bibfnamefont {E.~N.}\ \bibnamefont {Tsiok}}, \bibinfo
  {author} {\bibfnamefont {V.~V.}\ \bibnamefont {Brazhkin}}, \ and\ \bibinfo
  {author} {\bibfnamefont {K.}~\bibnamefont {Trachenko}},\ }\href {\doibase
  10.1088/0953-8984/28/43/43LT01} {\bibfield  {journal} {\bibinfo  {journal}
  {J. Phys.: Condens. Matter.}\ }\textbf {\bibinfo {volume} {28}},\ \bibinfo
  {pages} {43LT01} (\bibinfo {year} {2016})}\BibitemShut {NoStop}%
\bibitem [{\citenamefont {Fomin}\ \emph
  {et~al.}(2015{\natexlab{b}})\citenamefont {Fomin}, \citenamefont {Ryzhov},
  \citenamefont {Tsiok},\ and\ \citenamefont {Brazhkin}}]{Fomin2015}%
  \BibitemOpen
  \bibfield  {author} {\bibinfo {author} {\bibfnamefont {D.}~\bibnamefont
  {Fomin}}, \bibinfo {author} {\bibfnamefont {V.~N.}\ \bibnamefont {Ryzhov}},
  \bibinfo {author} {\bibfnamefont {E.~N.}\ \bibnamefont {Tsiok}}, \ and\
  \bibinfo {author} {\bibfnamefont {V.~V.}\ \bibnamefont {Brazhkin}},\ }\href
  {\doibase 10.1038/srep14234} {\bibfield  {journal} {\bibinfo  {journal} {Sci.
  Rep.}\ ,\ \bibinfo {pages} {14234}} (\bibinfo {year}
  {2015}{\natexlab{b}})}\BibitemShut {NoStop}%
\bibitem [{\citenamefont {Fomin}\ \emph {et~al.}(2014)\citenamefont {Fomin},
  \citenamefont {Ryzhov}, \citenamefont {Tsiok}, \citenamefont {Brazhkin},\
  and\ \citenamefont {Trachenko}}]{Fomin2014}%
  \BibitemOpen
  \bibfield  {author} {\bibinfo {author} {\bibfnamefont {Y.~D.}\ \bibnamefont
  {Fomin}}, \bibinfo {author} {\bibfnamefont {V.~N.}\ \bibnamefont {Ryzhov}},
  \bibinfo {author} {\bibfnamefont {E.~N.}\ \bibnamefont {Tsiok}}, \bibinfo
  {author} {\bibfnamefont {V.~V.}\ \bibnamefont {Brazhkin}}, \ and\ \bibinfo
  {author} {\bibfnamefont {K.}~\bibnamefont {Trachenko}},\ }\href {\doibase
  10.1038/srep07194} {\bibfield  {journal} {\bibinfo  {journal} {Sci. Rep.}\
  }\textbf {\bibinfo {volume} {4}},\ \bibinfo {pages} {7194} (\bibinfo {year}
  {2014})}\BibitemShut {NoStop}%
\bibitem [{\citenamefont {Yang}\ \emph {et~al.}(2015)\citenamefont {Yang},
  \citenamefont {Brazhkin}, \citenamefont {Dove},\ and\ \citenamefont
  {Trachenko}}]{Yang2015}%
  \BibitemOpen
  \bibfield  {author} {\bibinfo {author} {\bibfnamefont {C.}~\bibnamefont
  {Yang}}, \bibinfo {author} {\bibfnamefont {V.~V.}\ \bibnamefont {Brazhkin}},
  \bibinfo {author} {\bibfnamefont {M.~T.}\ \bibnamefont {Dove}}, \ and\
  \bibinfo {author} {\bibfnamefont {K.}~\bibnamefont {Trachenko}},\ }\href
  {\doibase 10.1103/PhysRevE.91.012112} {\bibfield  {journal} {\bibinfo
  {journal} {Phys. Rev. E}\ }\textbf {\bibinfo {volume} {91}},\ \bibinfo
  {pages} {012112} (\bibinfo {year} {2015})}\BibitemShut {NoStop}%
\bibitem [{\citenamefont {Bolmatov}\ \emph {et~al.}(2013)\citenamefont
  {Bolmatov}, \citenamefont {Brazhkin}, \citenamefont {Fomin}, \citenamefont
  {Ryzhov},\ and\ \citenamefont {Trachenko}}]{Bolmatov2013}%
  \BibitemOpen
  \bibfield  {author} {\bibinfo {author} {\bibfnamefont {D.}~\bibnamefont
  {Bolmatov}}, \bibinfo {author} {\bibfnamefont {V.~V.}\ \bibnamefont
  {Brazhkin}}, \bibinfo {author} {\bibfnamefont {Y.~D.}\ \bibnamefont {Fomin}},
  \bibinfo {author} {\bibfnamefont {V.~N.}\ \bibnamefont {Ryzhov}}, \ and\
  \bibinfo {author} {\bibfnamefont {K.}~\bibnamefont {Trachenko}},\ }\href
  {\doibase 10.1063/1.4844135} {\bibfield  {journal} {\bibinfo  {journal} {J.
  Chem. Phys.}\ }\textbf {\bibinfo {volume} {139}} (\bibinfo {year} {2013}),\
  10.1063/1.4844135}\BibitemShut {NoStop}%
\bibitem [{\citenamefont {Soper}(2000)}]{Soper2000}%
  \BibitemOpen
  \bibfield  {author} {\bibinfo {author} {\bibfnamefont {A.~K.}\ \bibnamefont
  {Soper}},\ }\href {\doibase 10.1016/S0301-0104(00)00179-8} {\bibfield
  {journal} {\bibinfo  {journal} {Chem. Phys.}\ }\textbf {\bibinfo {volume}
  {258}},\ \bibinfo {pages} {121} (\bibinfo {year} {2000})}\BibitemShut
  {NoStop}%
\bibitem [{\citenamefont {Santoro}\ and\ \citenamefont
  {Gorelli}(2008)}]{Santoro2008}%
  \BibitemOpen
  \bibfield  {author} {\bibinfo {author} {\bibfnamefont {M.}~\bibnamefont
  {Santoro}}\ and\ \bibinfo {author} {\bibfnamefont {F.}~\bibnamefont
  {Gorelli}},\ }\href {\doibase 10.1103/PhysRevB.77.212103} {\bibfield
  {journal} {\bibinfo  {journal} {Phys. Rev. B}\ }\textbf {\bibinfo {volume}
  {77}},\ \bibinfo {pages} {212103} (\bibinfo {year} {2008})}\BibitemShut
  {NoStop}%
\bibitem [{\citenamefont {Bolmatov}\ \emph {et~al.}(2015)\citenamefont
  {Bolmatov}, \citenamefont {Zhernenkov}, \citenamefont {Zav}, \citenamefont
  {Tkachev}, \citenamefont {Cunsolo},\ and\ \citenamefont
  {Cai}}]{Bolmatov2015}%
  \BibitemOpen
  \bibfield  {author} {\bibinfo {author} {\bibfnamefont {D.}~\bibnamefont
  {Bolmatov}}, \bibinfo {author} {\bibfnamefont {M.}~\bibnamefont
  {Zhernenkov}}, \bibinfo {author} {\bibfnamefont {D.}~\bibnamefont {Zav}},
  \bibinfo {author} {\bibfnamefont {S.~N.}\ \bibnamefont {Tkachev}}, \bibinfo
  {author} {\bibfnamefont {A.}~\bibnamefont {Cunsolo}}, \ and\ \bibinfo
  {author} {\bibfnamefont {Y.~Q.}\ \bibnamefont {Cai}},\ }\href {\doibase
  10.1038/srep15850} {\bibfield  {journal} {\bibinfo  {journal} {Sci. Rep.}\
  }\textbf {\bibinfo {volume} {5}},\ \bibinfo {pages} {15850} (\bibinfo {year}
  {2015})}\BibitemShut {NoStop}%
\bibitem [{\citenamefont {Brazhkin}\ and\ \citenamefont
  {Proctor}(2016)}]{comment}%
  \BibitemOpen
  \bibfield  {author} {\bibinfo {author} {\bibfnamefont {V.}~\bibnamefont
  {Brazhkin}}\ and\ \bibinfo {author} {\bibfnamefont {J.~E.}\ \bibnamefont
  {Proctor}},\ }\href@noop {} {\  (\bibinfo {year} {2016})}\BibitemShut
  {NoStop}%
\bibitem [{\citenamefont {Rivers}\ \emph {et~al.}(2008)\citenamefont {Rivers},
  \citenamefont {Prakapenka}, \citenamefont {Kubo}, \citenamefont {Pullins},
  \citenamefont {Holl},\ and\ \citenamefont {Jacobsen}}]{Rivers2008}%
  \BibitemOpen
  \bibfield  {author} {\bibinfo {author} {\bibfnamefont {M.}~\bibnamefont
  {Rivers}}, \bibinfo {author} {\bibfnamefont {V.}~\bibnamefont {Prakapenka}},
  \bibinfo {author} {\bibfnamefont {A.}~\bibnamefont {Kubo}}, \bibinfo {author}
  {\bibfnamefont {C.}~\bibnamefont {Pullins}}, \bibinfo {author} {\bibfnamefont
  {C.}~\bibnamefont {Holl}}, \ and\ \bibinfo {author} {\bibfnamefont
  {S.}~\bibnamefont {Jacobsen}},\ }\href {\doibase 10.1080/08957950802333593}
  {\bibfield  {journal} {\bibinfo  {journal} {High Press. Res.}\ }\textbf
  {\bibinfo {volume} {28}},\ \bibinfo {pages} {273} (\bibinfo {year}
  {2008})}\BibitemShut {NoStop}%
\bibitem [{\citenamefont {Mao}\ \emph {et~al.}(1986)\citenamefont {Mao},
  \citenamefont {Xu},\ and\ \citenamefont {Bell}}]{Mao1986}%
  \BibitemOpen
  \bibfield  {author} {\bibinfo {author} {\bibfnamefont {H.~K.}\ \bibnamefont
  {Mao}}, \bibinfo {author} {\bibfnamefont {J.}~\bibnamefont {Xu}}, \ and\
  \bibinfo {author} {\bibfnamefont {P.~M.}\ \bibnamefont {Bell}},\ }\href
  {\doibase 10.1029/JB091iB05p04673} {\bibfield  {journal} {\bibinfo  {journal}
  {J. Geophys. Res.}\ }\textbf {\bibinfo {volume} {91}},\ \bibinfo {pages}
  {4673} (\bibinfo {year} {1986})}\BibitemShut {NoStop}%
\bibitem [{\citenamefont {Weck}\ \emph {et~al.}(2013)\citenamefont {Weck},
  \citenamefont {Garbarino}, \citenamefont {Ninet}, \citenamefont {Spaulding},
  \citenamefont {Datchi}, \citenamefont {Loubeyre},\ and\ \citenamefont
  {Mezouar}}]{Weck2013}%
  \BibitemOpen
  \bibfield  {author} {\bibinfo {author} {\bibfnamefont {G.}~\bibnamefont
  {Weck}}, \bibinfo {author} {\bibfnamefont {G.}~\bibnamefont {Garbarino}},
  \bibinfo {author} {\bibfnamefont {S.}~\bibnamefont {Ninet}}, \bibinfo
  {author} {\bibfnamefont {D.}~\bibnamefont {Spaulding}}, \bibinfo {author}
  {\bibfnamefont {F.}~\bibnamefont {Datchi}}, \bibinfo {author} {\bibfnamefont
  {P.}~\bibnamefont {Loubeyre}}, \ and\ \bibinfo {author} {\bibfnamefont
  {M.}~\bibnamefont {Mezouar}},\ }\href {\doibase 10.1063/1.4807753} {\bibfield
   {journal} {\bibinfo  {journal} {Rev. Sci. Instrum.}\ }\textbf {\bibinfo
  {volume} {84}},\ \bibinfo {pages} {063901} (\bibinfo {year}
  {2013})}\BibitemShut {NoStop}%
\bibitem [{\citenamefont {Prescher}\ and\ \citenamefont
  {Prakapenka}(2015)}]{Prescher2015a}%
  \BibitemOpen
  \bibfield  {author} {\bibinfo {author} {\bibfnamefont {C.}~\bibnamefont
  {Prescher}}\ and\ \bibinfo {author} {\bibfnamefont {V.~B.}\ \bibnamefont
  {Prakapenka}},\ }\href {\doibase 10.1080/08957959.2015.1059835} {\bibfield
  {journal} {\bibinfo  {journal} {High Press. Res.}\ }\textbf {\bibinfo
  {volume} {35}},\ \bibinfo {pages} {223} (\bibinfo {year} {2015})}\BibitemShut
  {NoStop}%
\bibitem [{\citenamefont {Eggert}\ \emph {et~al.}(2002)\citenamefont {Eggert},
  \citenamefont {Weck}, \citenamefont {Loubeyre},\ and\ \citenamefont
  {Mezouar}}]{Eggert2002}%
  \BibitemOpen
  \bibfield  {author} {\bibinfo {author} {\bibfnamefont {J.}~\bibnamefont
  {Eggert}}, \bibinfo {author} {\bibfnamefont {G.}~\bibnamefont {Weck}},
  \bibinfo {author} {\bibfnamefont {P.}~\bibnamefont {Loubeyre}}, \ and\
  \bibinfo {author} {\bibfnamefont {M.}~\bibnamefont {Mezouar}},\ }\href
  {\doibase 10.1103/PhysRevB.65.174105} {\bibfield  {journal} {\bibinfo
  {journal} {Phys. Rev. B}\ }\textbf {\bibinfo {volume} {65}},\ \bibinfo
  {pages} {174105} (\bibinfo {year} {2002})}\BibitemShut {NoStop}%
\bibitem [{\citenamefont {Oh}(2013)}]{Oh2013}%
  \BibitemOpen
  \bibfield  {author} {\bibinfo {author} {\bibfnamefont {S.~K.}\ \bibnamefont
  {Oh}},\ }\href {\doibase 10.1155/2013/828620} {\bibfield  {journal} {\bibinfo
   {journal} {J. Thermodyn.}\ ,\ \bibinfo {pages} {82860}} (\bibinfo {year}
  {2013})}\BibitemShut {NoStop}%
\bibitem [{\citenamefont {Plimpton}(1995)}]{lammps}%
  \BibitemOpen
  \bibfield  {author} {\bibinfo {author} {\bibfnamefont {S.}~\bibnamefont
  {Plimpton}},\ }\href {\doibase 10.1006/jcph.1995.1039} {\bibfield  {journal}
  {\bibinfo  {journal} {J. Comput. Phys.}\ }\textbf {\bibinfo {volume} {117}},\
  \bibinfo {pages} {1} (\bibinfo {year} {1995})}\BibitemShut {NoStop}%
\bibitem [{\citenamefont {Frenkel}\ and\ \citenamefont
  {Smit}(2002)}]{Frenkel2002}%
  \BibitemOpen
  \bibfield  {author} {\bibinfo {author} {\bibfnamefont {D.}~\bibnamefont
  {Frenkel}}\ and\ \bibinfo {author} {\bibfnamefont {B.}~\bibnamefont {Smit}},\
  }\href@noop {} {\emph {\bibinfo {title} {{Understanding Molecular
  Simulation}}}}\ (\bibinfo  {publisher} {Academic Press},\ \bibinfo {address}
  {London},\ \bibinfo {year} {2002})\ p.\ \bibinfo {pages} {638}\BibitemShut
  {NoStop}%
\bibitem [{\citenamefont {Lemmon}\ \emph {et~al.}(2016)\citenamefont {Lemmon},
  \citenamefont {McLinden},\ and\ \citenamefont {Friend}}]{Lemmon}%
  \BibitemOpen
  \bibfield  {author} {\bibinfo {author} {\bibfnamefont {E.}~\bibnamefont
  {Lemmon}}, \bibinfo {author} {\bibfnamefont {M.}~\bibnamefont {McLinden}}, \
  and\ \bibinfo {author} {\bibfnamefont {D.~G.}\ \bibnamefont {Friend}},\ }in\
  \href@noop {} {\emph {\bibinfo {booktitle} {NIST Chemistry WebBook, NIST
  Standard Reference Database Number 69}}},\ \bibinfo {editor} {edited by\
  \bibinfo {editor} {\bibfnamefont {P.}~\bibnamefont {Linstrom}}\ and\ \bibinfo
  {editor} {\bibfnamefont {W.}~\bibnamefont {Mallard}}}\ (\bibinfo  {publisher}
  {National Institute of Standards and Technology},\ \bibinfo {address}
  {Gaithersburg MD, 20899},\ \bibinfo {year} {2016})\BibitemShut {NoStop}%
\bibitem [{\citenamefont {Wang}\ \emph {et~al.}(2017)\citenamefont {Wang},
  \citenamefont {Yang}, \citenamefont {Dove}, \citenamefont {Fomin},
  \citenamefont {Brazhkin},\ and\ \citenamefont {Trachenko}}]{Wang2017}%
  \BibitemOpen
  \bibfield  {author} {\bibinfo {author} {\bibfnamefont {L.}~\bibnamefont
  {Wang}}, \bibinfo {author} {\bibfnamefont {C.}~\bibnamefont {Yang}}, \bibinfo
  {author} {\bibfnamefont {M.~T.}\ \bibnamefont {Dove}}, \bibinfo {author}
  {\bibfnamefont {Y.~D.}\ \bibnamefont {Fomin}}, \bibinfo {author}
  {\bibfnamefont {V.~V.}\ \bibnamefont {Brazhkin}}, \ and\ \bibinfo {author}
  {\bibfnamefont {K.}~\bibnamefont {Trachenko}},\ }\href {\doibase
  10.1103/PhysRevE.95.032116} {\bibfield  {journal} {\bibinfo  {journal} {Phys.
  Rev. E.}\ }\textbf {\bibinfo {volume} {95}},\ \bibinfo {pages} {032116}
  (\bibinfo {year} {2017})}\BibitemShut {NoStop}%
\bibitem [{\citenamefont {Trachenko}\ and\ \citenamefont
  {Brazhkin}(2016)}]{Trachenko2016}%
  \BibitemOpen
  \bibfield  {author} {\bibinfo {author} {\bibfnamefont {K.}~\bibnamefont
  {Trachenko}}\ and\ \bibinfo {author} {\bibfnamefont {V.~V.}\ \bibnamefont
  {Brazhkin}},\ }\href {\doibase 10.1088/0034-4885/79/1/016502} {\bibfield
  {journal} {\bibinfo  {journal} {Rep. Prog. Phys.}\ }\textbf {\bibinfo
  {volume} {79}},\ \bibinfo {pages} {016502} (\bibinfo {year}
  {2016})}\BibitemShut {NoStop}%
\bibitem [{\citenamefont {Hansen}\ and\ \citenamefont
  {McDonald}(2013)}]{Hansen2013}%
  \BibitemOpen
  \bibfield  {author} {\bibinfo {author} {\bibfnamefont {J.~P.}\ \bibnamefont
  {Hansen}}\ and\ \bibinfo {author} {\bibfnamefont {I.~R.}\ \bibnamefont
  {McDonald}},\ }\href@noop {} {\emph {\bibinfo {title} {{Theory of Simple
  Liquids}}}},\ \bibinfo {edition} {4th}\ ed.\ (\bibinfo  {publisher} {Academic
  Press},\ \bibinfo {address} {San Diego},\ \bibinfo {year} {2013})\ p.\
  \bibinfo {pages} {636}\BibitemShut {NoStop}%
\bibitem [{\citenamefont {Rapaport}(1995)}]{Rapaport1995}%
  \BibitemOpen
  \bibfield  {author} {\bibinfo {author} {\bibfnamefont {D.~C.}\ \bibnamefont
  {Rapaport}},\ }\href@noop {} {\emph {\bibinfo {title} {{The Art of Molecular
  Dynamics Simulations}}}}\ (\bibinfo  {publisher} {Cambridge University
  Press},\ \bibinfo {address} {Cambridge},\ \bibinfo {year} {1995})\ p.\
  \bibinfo {pages} {564}\BibitemShut {NoStop}%
\bibitem [{\citenamefont {Brazhkin}\ \emph {et~al.}(2002)\citenamefont
  {Brazhkin}, \citenamefont {Buldyrev}, \citenamefont {Ryzhov},\ and\
  \citenamefont {Stanley}}]{Brazhkin2002}%
  \BibitemOpen
  \bibinfo {editor} {\bibfnamefont {V.~V.}\ \bibnamefont {Brazhkin}}, \bibinfo
  {editor} {\bibfnamefont {S.~V.}\ \bibnamefont {Buldyrev}}, \bibinfo {editor}
  {\bibfnamefont {V.~N.}\ \bibnamefont {Ryzhov}}, \ and\ \bibinfo {editor}
  {\bibfnamefont {H.~E.}\ \bibnamefont {Stanley}},\ eds.,\ \href@noop {} {\emph
  {\bibinfo {title} {{New Kinds of Phase Transitions: Transformations in
  Disordered Substances, Proceeding of NATO Advanced Research Workshop, Volga
  River}}}}\ (\bibinfo  {publisher} {Kluwer Academic Publishers},\ \bibinfo
  {address} {Dordrecht},\ \bibinfo {year} {2002})\BibitemShut {NoStop}%
\bibitem [{\citenamefont {Kiran}\ \emph {et~al.}(2000)\citenamefont {Kiran},
  \citenamefont {Debenedetti},\ and\ \citenamefont {Peters}}]{Kiran2000}%
  \BibitemOpen
  \bibfield  {author} {\bibinfo {author} {\bibfnamefont {E.}~\bibnamefont
  {Kiran}}, \bibinfo {author} {\bibfnamefont {P.}~\bibnamefont {Debenedetti}},
  \ and\ \bibinfo {author} {\bibfnamefont {C.~J.}\ \bibnamefont {Peters}},\
  }\href@noop {} {\emph {\bibinfo {title} {{Supercritical Fluids}}}}\ (\bibinfo
   {publisher} {Springer-Science+Business Media, BV},\ \bibinfo {address}
  {Dordrecht},\ \bibinfo {year} {2000})\ p.\ \bibinfo {pages} {596}\BibitemShut
  {NoStop}%
\end{thebibliography}%

\end{document}